\documentclass[apj,twocolumn]{emulateapj}
\usepackage{apjfonts,epsfig,mathptmx}
\usepackage[dvipdfm]{hyperref}
\shortauthors{Ma et al.}
\shorttitle{Physical Properties of cSFGs}

\begin{document}

\title{Physical properties of compact star-forming galaxies at $z\sim2-3$}

\author{
Guanwen Fang\altaffilmark{1,2},
Zhongyang Ma\altaffilmark{2,3*},
Xu Kong\altaffilmark{2,3},
Lulu Fan\altaffilmark{4}
}

\altaffiltext{1}{Institute for Astronomy and History of Science and Technology, Dali University, Dali 671003, China}
\altaffiltext{2}{Key Laboratory for Research in Galaxies and Cosmology, CAS, Hefei 230026, China}
\altaffiltext{3}{Department of Astronomy, University of Science and Technology of China, Hefei 230026, China}
\altaffiltext{4}{Shandong Provincial Key Laboratory of Optical Astronomy and Solar-Terrestrial Environment, Institute of Space Sciences, Shandong University, Weihai 264209, China}
\altaffiltext{*}{Zhongyang Ma and Guanwen Fang contributed equally to this work.}

\email{wen@mail.ustc.edu.cn, xkong@ustc.edu.cn}

\begin{abstract}

We present a study on the physical properties of compact star-forming galaxies (cSFGs) with $M_{*}\geq10^{10}M_{\odot}$ and $2\leq z\leq3$ in the COSMOS and GOODS-S fields. We find that massive cSFGs have a comoving number density of $(1.0\pm0.1)\times10^{-4}~{\rm Mpc}^{-3}$.
The cSFGs are distributed at nearly the same locus on the main sequence as extended star-forming galaxies (eSFGs) and dominate the high-mass end. On the rest-frame $U-V$ vs. $V-J$ and $U-B$ vs. $M_{\rm B}$ diagrams, cSFGs are mainly distributed at the middle of eSFGs and compact quiescent galaxies (cQGs) in all colors, but are more inclined to ``red sequence'' than ``green valley'' galaxies. We also find that cSFGs have distributions similar to cQGs on the nonparametric morphology diagrams. The cQGs and cSFGs have larger $Gini$ and smaller $M_{20}$, while eSFGs have the reverse. About one-third of cSFGs show signatures of postmergers, and almost none of them can be recognized as disks. Moreover, those visually extended cSFGs all have lower $Gini$ coefficients ($Gini<0.4$), indicating that
the $Gini$ coefficient could be used to clean out noncompact galaxies in a sample of candidate cSFGs.
The X-ray-detected counterparts are more frequent among cSFGs than that in eSFGs and cQGs,
implying that cSFGs have previously experienced violent gas-rich interactions(such as major mergers or disk instabilities),
which could trigger both star formation and black hole growth in an active phase.

\end{abstract}

\section{INTRODUCTION}

Galaxies in local universe present a bimodal distribution in colors, as introduced by previous literatures: the ``red sequence'' which mainly composed by quiescent galaxies (QGs) with older stellar populations and a certain amount of dusty star-forming galaxies (SFGs) \citep{Blanton2009}, and the ``blue cloud'' which composed by young active star-forming galaxies with extended structures \citep{Kauffmann2003,Baldry2004}. And the bimodal distribution in colors of galaxies is confirmed to be existed already at $z\sim2-3$ \citep{Faber2007,Ilbert2010,Brammer2011}.

The star formation status correlates a wide range of physical properties of galaxy, in other words, it reflects that galaxies experienced different physical states in evolution. At $z\sim2$, most of QGs are visually compact, round and centrally concentrated, with no extended structure, they generally have larger S\'{e}rsic index $n$ and $Gini$ coefficient, smaller effective radius $r_{\rm e}$ and moment index $M_{20}$. While SFGs are opposite to QGs: they are visually extended, and have a disk-like or irregular morphologies, sometimes show interaction features such as tidal arms, and they are commonly larger than QGs with similar masses \citep{Bell2012,Fang2012,vdwel2012,vdwel2014,Wang2012,Lee2013}. Recent works have shown that there is an evolutionary connection between these two populations \citep{Buitrago2013,Fang2012,Fan2013,Patel2013,Toft2014}. The evolutionary scenario is that through wet merging process at $z>2$, the extended SFGs (eSFGs) lose their angular momentum, in the mean time, there are large amounts of stars formed, then they left behind a compact relics: the compact QGs (cQGs), or the bulge of quiescent disk galaxies \citep{Naab2007,Hopkins2008,Oser2010,Fan2013,vanDokkum2014,Stringer2015,Wellons2015}. The fraction of compact or ultra-compact QGs is higher at $z>2$, while most of QGs formed at $z<2$ have sizes comparable to those of local counterparts of the same masses \citep{Cassata2013}. If the evolutionary scenario for QGs described above is real, a co-existing population of galaxies which is considered to be an exhibit of evolutionary connection between extend SFGs (eSFGs) and compact QGs (cQGs) are expected to be observed at $z>2$ \citep{Barro2013,Barro2015}.

Recently, people find an interesting population of galaxies at $z\sim2-4$ \citep{Cava2010,Wuyts2011b,Whitaker2012a,Kaviraj2013,Stefanon2013,Williams2014}, their structures are very compact, nearly the same as QGs, but their star formation status are still very active, almost as active as normal SFGs \citep{Barro2013,Barro2014a,Barro2014b}. This is the population of galaxies which will be particularly analyzed in this paper: the compact star-forming galaxies (cSFGs). The cSFG is a transitional type of galaxy evolution between extend SFG (eSFG) and compact QG (cQG) at $z>2$, and both mergers and disk instabilities are able to shrink galaxies from eSFGs to cSFGs \citep{Barro2014a}, then when star formation activities are totally quenched, they become cQGs. If it can be confirmed that cSFGs are truly the direct progenitors of cQGs at $z>2$, we propose these evolutionary tracks of nearby QGs as follows: at $z>2$, cSFGs rapidly quenched into cQGs which become the compact core parts of local QGs later, then they enlarge their outer sizes through non-dissipative dry mergers to become normal QGs in local universe. Dry merger can not effectively form stars, but can significantly enlarge the size of galaxy \citep{Trujillo2007,Buitrago2008,Fan2013}. Another possible formation path is that eSFGs directly form extend QGs (eQGs) at $z<2$ by AGN or supernova feed back, they have not experienced a compact status \citep{Barro2013}.

With a sample of candidate progenitors of $z\sim 1.6$ cQGs among compact Lyman-break galaxies (LBGs) at $z\sim 3$,
\cite{Williams2014} compare their properties to those of non-candidate normal SFGs, and find that the average far-UV
spectral energy distribution (SED) of the cSFGs is redder than that of the normal SFGs, but the optical and mid-IR
SED are the same. Concurrently, they suggest the cQGs are formed primarily through the quenching of cSFGs whose in
situ star formation is driven by cold accretion from the intergalactic medium via violent disk instability and
cold mode accretion. To study the nature of the quenching mechanism affecting compact galaxies, \cite{Williams2015} present
a comparative analysis of rest-frame UV spectroscopy of 12 cSFGs presented in \cite{Williams2014}, compared with
the properties of 20 normal SFGs at the same redshift. The findings show that the faster bulk motions, broader
spread of gas velocity, and Ly$\alpha$ properties of cSFGs are consistent with their interstellar medium being
subject to more energetic feedback than normal SFGs.
\cite{Barro2014a} analyzed the star-forming and structural properties of 45 massive cSFGs at $2<z<3$ in GOODS-S field based on Cosmic Assembly Near-infrared Deep Extragalactic Legacy Survey (CANDELS) \citep{Grogin2011,Koekemoer2011} and 3D-HST \citep{Brammer2012} data, to explore whether they are natural progenitors of compact quiescent galaxies at $z\sim2$. They show that galaxies become more compact before they lose their gas and dust, and most of cSFGs are heavily obscured, 47\% of them host an X-ray-bright AGN, and 65\% of them distribute at red sequence.

Since the samples of previous research programs of are commonly small, it still remains many uncertainties
on the physical properties of massive cSFGs at $2<z<3$ (owing to the cosmic variance). In order to have a thorough understanding of this
kind of galaxies, in this work we make a statistical analysis on the physical properties of cSFGs selected from COSMOS field and the 45 cSFGs in GOODS-S field \citep{Barro2014a}, based on the new released 3D-HST \citep{Skelton2014} high-quality multi-wavelength photometric data and CANDELS \citep{Grogin2011,Koekemoer2011} imaging data. For a sample of 104 cSFGs (COSMOS+GOODS-S) with $M_{*}\geq10^{10}M_{\odot}$ at $2<z<3$,
we perform nonparametric measures of galaxy morphology for the first time.
In the meantime, we also analyze their co-moving number density, rest-frame colors, visual morphologies, distributions on stellar population and structural parameters, and AGN fractions of them, and discuss the difference of physical properties among eSFGs, cSFGs and cQGs.

The paper is organized as follows. We introduce the CANDELS and 3D-HST data in Section 2, the selection of our cSFG sample in Section 3. We show the results of the physical properties of cSFGs in Section 4 and conclude our results in Section 5. Throughout this paper, we assume an $\Omega_\mathrm{M}=0.3, \Omega_{\Lambda}=0.7$ and $H_{0}=70$ km s$^{-1}$ Mpc$^{-1}$ cosmology, a \cite{Chabrier2003} initial mass function (IMF), and all magnitudes and colors are given in AB system unless stated otherwise.

\section{OBSERVATIONS AND DATA}

Our study is based on a sample of high-redshift massive galaxies which built from 3D-HST \citep{Skelton2014} and CANDELS \citep{Grogin2011,Koekemoer2011} data. The 3D-HST and CANDELS programs have provided WFC3 and ACS spectroscopy and photometry over $\sim900$ arcmin$^2$ in five fields: AEGIS, COSMOS, GOODS-North, GOODS-South, and the UKIDSS UDS field. All these fields have a wealth of publicly available imaging datasets in addition to the HST data, which makes it possible to construct the SEDs of objects over a wide wavelength range \citep{Skelton2014}.

The derived data products are also provided by \cite{Skelton2014}, stellar masses and other stellar population parameters were determined by {\tt FAST} \citep{Kriek2009} code, photometric redshifts and rest-frame colors were derived using the {\tt EAZY} \citep{Brammer2008} code. 3D-HST is a spectroscopic survey with the WFC3 and ACS grisms, we will use the spectroscopic redshifts if the objects have, otherwise we use their derived photometric redshifts. The qualities of the derived photometric redshifts are very good, the normalized median absolute deviations $\sigma_\mathrm{NMAD}$ of photometric redshifts versus spectroscopic redshifts for COSMOS are 0.007.

Structural parameters of galaxy such as S\'{e}rsic index $n$, effective radii $r_{\rm e}$, axis ratio $q$ come from the catalog of \cite{vdwel2014}, these parameters are measured from CANDELS WFC3 $H$-band images with {\tt GALFIT} \citep{Peng2002}. Our morphology analysis was enabled by the $HST$/WFC3 NIR imaging from the CANDELS \citep{Grogin2011,Koekemoer2011}. We measure the nonparametric morphologies, such as $Gini$ and $M_{20}$ of galaxies using the program developed by \cite{Abraham2007}. We match our sample with the CANDELS F160W ($H$-band) image, which corresponds to the rest-frame optical morphologies of galaxies distribute at $z\sim2-3$.

The SFRs of galaxies in COSMOS field were matched from UltraVISTA catalog \citep{Muzzin2013}, these SFRs are converted from $L_{2800}$ and $L_\mathrm{IR}$, using the conversion factors SFR$_\mathrm{UV,uncorr}=3.234\times10^{-10}L_{2800}$ and SFR$_\mathrm{IR}=0.98\times10^{-10}L_\mathrm{IR}$ from \cite{Kennicutt1998}. The total SFR of the galaxy can then be determined via SFR$_\mathrm{tot}$=SFR$_\mathrm{UV,uncorr}$+SFR$_\mathrm{IR}$.
The SFR is measured based on UltraVISTA redshift ($z_{\rm UltraVISTA}$) and $M_{\ast}$ is
measured based on 3D-HST redshift ($z_{\rm 3D-HST}$), there is difference between the two
redshifts, but it is small. From the comparison between $z_{\rm 3D-HST}$
and $z_{\rm UltraVISTA}$, we find that UltraVISTA photometric redshifts are in good agreement with the
3D-HST photometric redshifts, with an average $(z_{\rm 3D-HST}-z_{\rm UltraVISTA})/(1+z_{\rm UltraVISTA})=0.012$.
The normalized median absolute deviation, $\sigma_{\rm NMAD}$, for galaxies is  $\sigma_{\rm NMAD}= 0.022$.
Statistically, such a small difference would not introduce any systematic offsets to SFR (although the SFR of individual
galaxies is changed from using $z_{\rm UltraVISTA}$ to using $z_{\rm 3D-HST}$).

\section{SAMPLE SELECTION}

We select a sample of cSFGs at $2\leq z\leq3$, using the criteria defined by \cite{Barro2014a}. The criteria are conclude as follows:

\begin{equation}
 \mathrm{log}(M_{*}/r_{\rm e}^{1.5})\geq 10.45 \ M_{\odot}\ \mathrm{kpc}^{-1.5},
\end{equation}
\begin{equation}
 \mathrm{log}\mathrm{~sSFR}\geq -9.75\ \mathrm{yr^{-1}},
\end{equation}
\begin{equation}
 2\leq z \leq3,
\end{equation}
\begin{equation}
 M_{*}\geq10^{10}M_{\odot}.
\end{equation}

Equation (1) defines the compactness of a galaxy using a threshold in $pseudo$-stellar mass surface density $\Sigma_{1.5}$, log($M_{*}/r_\mathrm{e}^{1.5}$). Equation (2) defines the star formation activity level of a galaxy by limit its specific SFR (sSFR). Equation (3) and (4) imply we focus on the massive cSFGs at high-redshifts, for the purpose of comparing our sample to previous works, these redshift and mass criteria are consistent to \cite{Barro2014a}.

We only select object with the flag {\tt use\_phot}=1 in \cite{Skelton2014}, which means the object (1) not a star, or too faint to be recognized as a star or a galaxy. (2) not close to a bright star. (3) well-exposed in the F125W and F160W. (4) S/N$>$3 in F160W images. (5) ``non-catastrophic" photometric redshift and stellar population fits \citep{Skelton2014}. Figure 1 shows the motivation of these selection criteria: the mass-size relation for galaxies more massive than $10^{10}M_{\odot}$ at $2\leq z\leq3$. The subpopulations of compact QGs and SFGs, extend QGs and SFGs are plotted in red, green, orange, blue, respectively. Totally, we identify 59 cSFGs in COSMOS field (see Table 1). It is easy to find that the compactness-selected cSFGs follow a similar mass-size relation with compact quiescent galaxies, which follow a tight mass-size relation with a slope $\sim1.5$ that remain constant with redshifts \citep{Barro2014b}. We also combined the 45 cSFGs selected from GOODS-S \citep{Barro2014a} to our sample for the quantitative analysis on each physical parameters. For cSFGs from the GOODS-S field, their co-moving number density corresponds to
$(1.2\pm0.2)\times10^{-4}~{\rm Mpc}^{-3}$. The number density of cSFGs in our sample is $(0.9\pm0.2)\times10^{-4}~{\rm Mpc}^{-3}$
between $2<z<3$. It is consistent within the uncertainty with the volume density of the \citep{Barro2014a} sample. To reduce the influence of
cosmic variance on number density, we calculated the number density of cSFGs by combining cSFGs from the COSMOS and GOODS-S fields. We find
that massive ($M_{*}\geq10^{10}M_{\odot}$) cSFGs have a co-moving number density of $(1.0\pm0.1)\times10^{-4}~{\rm Mpc}^{-3}$ at $2<z<3$.

\begin{figure}
\includegraphics[width=0.95\columnwidth]{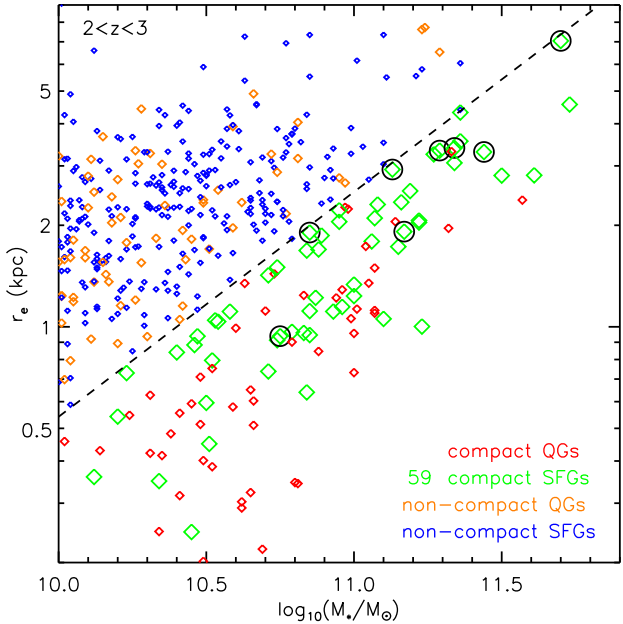}
\caption{Sample selection for cSFGs in COSMOS field at $2\leq z\leq3$. The dashed line represents the compactness criterion log($M_{*}/r_\mathrm{e}^{1.5}$)=10.45 $M_{\odot}$ kpc$^{-1.5}$ (below it are compact galaxies). SFGs are separated by the specific SFR (sSFR) $\rm log(sSFR)\geq-9.75$ $\rm yr^{-1}$ from QGs. The compact SFGs with $Gini<0.4$ are marked by black circles (please see Section 4.3 for more details).}
\end{figure}

\section{PHYSICAL PROPERTIES OF cSFGs}

In this section, we present the statistical results of some aspects of physical properties for eSFGs, cSFGs and cQGs.
Physical properties of galaxies, such as stellar mass, SFR, color, morphology, size, age, dust extinction, and fraction of AGNs, are derived on the basis of the present photometric data and $HST$ WFC3/H(F160W) imaging.

\subsection{The Star-forming Main Sequence}

We discuss the star-forming status of cSFGs starting from the main sequence, which represents the relation between stellar masses and SFRs of galaxies \citep{Daddi2007,Elbaz2007,Barro2013}. In Figure 2, we show the main sequence (MS) for eSFGs (blue), cSFGs (green) and cQGs (red) in COSMOS field at $2\leq z<2.5$ in the left panel and $2.5\leq z\leq 3$ in the right panel, and we also plot the cSFGs from GOODS-S \citep{Barro2014a} in cyan squares.
We mark the QG selecting threshold $\rm log~sSFR<-9.75~yr^{-1}$ using red dotted lines. Combined with the cSFGs in COSMOS (green) and GOODS-S (cyan), we fit the main sequence relation of them in dark green solid lines, and the 1$\sigma$ dispersion of these fits are shown by dark green dashed lines, we give the fit $y=-7.13+0.83x$, $\sigma=0.47$ for 63 cSFGs distribute at $2\leq z<2.5$, and give the fit $y=-9.84+1.09x$, $\sigma=0.50$ for 41 cSFGs at $2.5\leq z\leq3$. Figure 2 also shows that the relation between stellar mass and SFR in different redshift bins for eSFGs (blue lines). At fixed stellar mass, star-forming galaxies were much more active on average, compared to quiescent galaxies. Moreover, almost all of massive cSFGs follow the MS, implying that these cSFGs share similar stellar population properties.

Figure 2 clearly shows that cSFGs distribute at nearly the same locus on the main sequence as eSFGs, but cSFGs have slightly lower SFRs than eSFGs, and they dominate the high mass end of the main sequence, which could be possibly explained by that some of them are on the way to quiescent population in the evolutionary path. The cSFGs are more massive than eSFGs, however, they remain the highly active star-forming status into a compact phase. If cSFGs are the descendants of higher redshift eSFGs, they must have experienced violent interactions in a gas-rich environment, and enough amount of gas should be remained in the cSFGs to contribute to the high level of star formation activity. However, we find no starbursts in our sample, which possibly suggest that the starburst phase is short-lived compared to star formation activity and it occurs prior to the compact phase.

\begin{figure*}
\center
\includegraphics[width=0.8\textwidth]{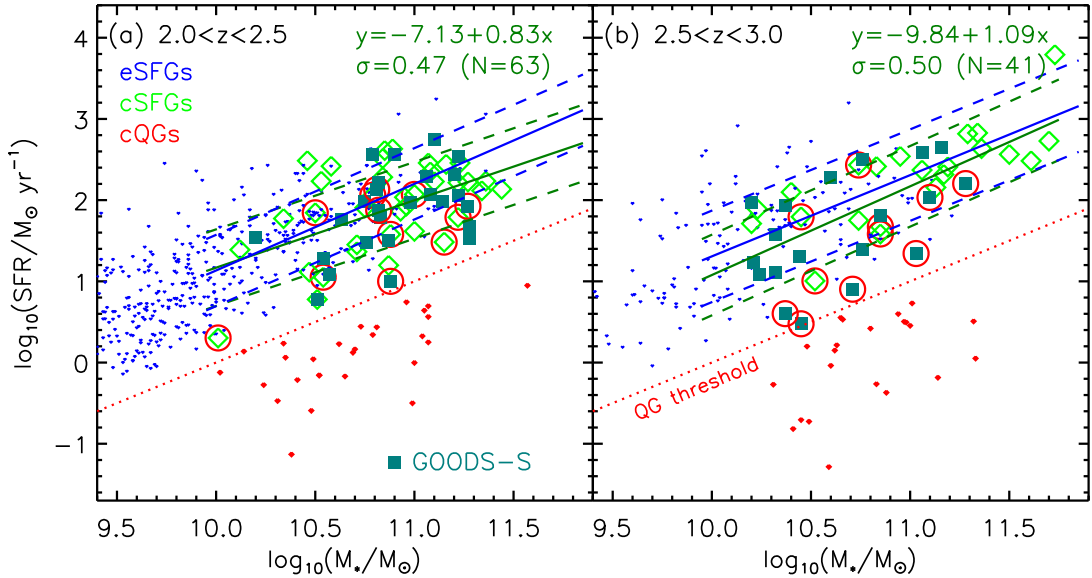}
\caption{The main sequence diagram for eSFGs (blue), cSFGs (green) and cQGs (red) in COSMOS field at $2\leq z<2.5$ (left panel) and $2.5\leq z\leq3$ (right panel). {\bf The cSFGs that fall into the $UVJ$ quiescent region} are marked with red circles (please see Section 4.2 for more details). cSFGs from GOODS-S are plotted in cyan squares.
The red dot lines represent the QG threshold $\rm log~sSFR<-9.75~yr^{-1}$. Combined with the cSFGs in COSMOS (green) and GOODS-S (cyan), we fit the main sequence relation of them in dark green solid lines, and the 1$\sigma$ dispersion of these fits are shown by dark green dashed lines. The blue solid line shows the MS relation in different redshift bins for eSFGs.}
\end{figure*}

\subsection{Color-color and Color-magnitude Diagram}

We explore the star-forming status and the extinction properties of cSFGs from the perspective of the distribution on rest-frame colors. The left panel of Figure 3 shows the distribution on rest-frame $U-V$ vs. $V-J$ ($UVJ$) diagram for eSFGs (blue), cSFGs (green) and cQGs (red) at $2\leq z\leq3$. And cSFGs from GOODS-S are plotted in cyan squares. The quiescent and star-forming separation lines are consistent with that of \cite{Williams2009} and \cite{Skelton2014}. The $UVJ$ diagram has been proved to be a successful method to distinguish the older quiescent and dusty star-forming populations for galaxies based on their SEDs. We find that most of cSFGs have redder $V-J$ colors, which means they have larger dust extinctions, and they are more close to the quiescent region than eSFGs. About 20\% of cSFGs located in the quiescent region, but they have bluer $V-J$ colors and are more close to the boundary than cQGs, these cSFGs are supposed to be close to become the quenched compact galaxies. The results of the distribution on $UVJ$ colors of cSFG are consistent with that on the ``main sequence'' diagram, indicating that cSFG is a possible intermediate type of galaxy formation and evolution.

\begin{figure*}
\center
\includegraphics[width=0.8\textwidth]{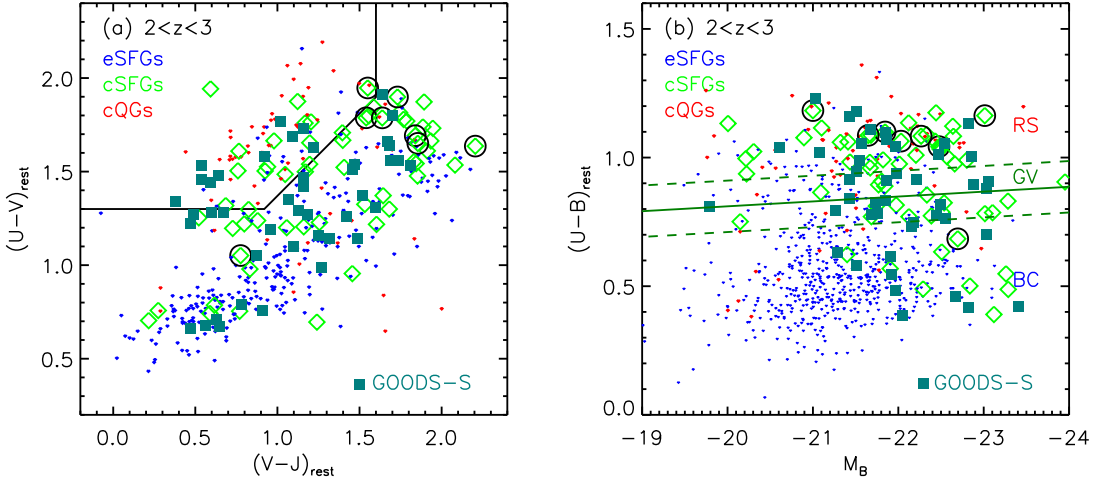}
\caption{Left panel: Distribution on rest-frame $U-V$ vs. $V-J$ ($UVJ$) diagram for eSFGs (blue), cSFGs (green) and cQGs (red) at $2\leq z\leq3$. And cSFGs from GOODS-S field are plotted in cyan squares. The quiescent and star-forming separation lines are consistent with that of Williams et al. (2009) and Skelton et al. (2014). Right panel: Distribution on rest-frame $U-B$ vs. $M_{\rm B}$ diagram for eSFGs (blue), cSFGs (green) and cQGs (red) at $2\leq z\leq3$. And cSFGs from GOODS-S field are also plotted in cyan squares. The dark green solid line and dashed lines represent the ``Green Valley'' (GV) which is defined as $y=-0.019(x+20.5)+0.82$ $(\pm0.1)$. The ``red sequence'' (RS) and ``blue cloud'' (BC) regions are also marked in this figure. The cSFGs of $Gini<0.4$ are marked by black circles.}
\end{figure*}

The ``Green Valley'' galaxy is supposed to be intermediate between the red and blue galaxy populations in terms of a series of physical properties, and most of them are compact disks and lack of mergers \citep{Mendez2011}. The cSFGs are also considered to be intermediate, so we are curious that whether cSFGs are ``Green Valley'' galaxies and how many overlaps between them. The right panel of Figure 3 shows the distribution on the rest-frame $U-B$ vs. $M_{\rm B}$ diagram for eSFGs (blue), cSFGs (green) and cQGs (red) at $2\leq z\leq3$. And cSFGs from GOODS-S are also plotted in cyan squares. The dark green solid line and dashed lines represent the ``Green Valley'' (GV) which is defined as $y=-0.019(x+20.5)+0.82$ $(\pm0.1)$, using the same method as \cite{Mendez2011} based on our sample. The ``red sequence'' (RS) and ``blue cloud'' (BC) regions are also marked in this figure. We find that the percentages of cSFGs distributing in the region of RS, GV and BC are 47\%, 33\% and 20\%, respectively. Most of (80\%) cSFGs located in RS and GV regions, while nearly all the eSFGs located in BC region and nearly all the cQGs located in RS region. We can also clearly find that the cSFGs distribute between eSFGs and cQGs in rest-frame $U-B$ color, but they dominate the high luminosity end. The result from the $U-B$ vs. $M_{\rm B}$ diagram also suggest the idea that cSFG is a possible transitional phase of galaxy
 formation and evolution, but the distribution on $U-B$ color of them are extended than ``Green Valley'' galaxies and are more inclined to ``red sequence''.

\subsection{Morphology}

Morphologies of galaxies correlate with a series of physical properties, and can provide direct information on the formation and evolution history of these objects. Owning to the observed optical light probes the rest-frame UV emission for objects at $z\sim2$, their apparent morphologies can easily be changed by patchy dust extinction. The rest-frame UV emission of galaxies mainly contributed by the hottest stars and can be heavily affected by dust extinction, therefore it is essential to study $2\leq z\leq3$ galaxies from F160W ($H-$band) images, which correspond to their rest-frame optical morphologies.

\subsubsection{Non-parametric Measurements}

To describe the morphological properties of galaxies in our sample, we have performed nonparametric measures of galaxy morphology, such as $Gini$ coefficient (the relative distribution of the galaxy pixel flux values) \citep{Abraham2003} and $M_{20}$ (the second-order moment of the brightest 20\% of the galaxy's flux) \citep{Lotz2004}, using the Morpheus-software developed by \cite{Abraham2007}. Morpheus is a collection of programs for automated morphological measurement
and classification. The code for calculating the morphological statistics has been modified to include new statistics and accommodate much larger input images \citep{Abraham2007}. In addition
to incorporating some relatively new parameters ($Gini$ coefficient and $M_{20}$), Morpheus also incorporates
improvements suggested by others for ways to better measure well-established
parameters such as $Asymmetry$ \citep{Conselice2000,Conselice2003,Conselice2005}.

As described in \cite{Lotz2004},
\begin{equation}
\label{eq:G} Gini = \frac{\sum^{N}_{l} (2l-N-1) |F_{l}|}{\overline{F}N(N-1)},
\end{equation}
where $N$ is the total number of pixels in a galaxy, and $\overline{F}$ is the
mean pixel flux of all $F_{l}$ (each pixel flux).
\begin{equation}
\label{eq:M20} M_{\rm 20} = {\rm log}(\frac{\sum^{k}_{l=1}M_{l}}{M_{\rm tot}}),
\end{equation}
${\rm where} \sum^{k}_{l=1}F_{l}=0.2F_{\rm tot} {~\rm and~} M_{\rm tot}=\sum^{N}_{l=1}M_{l}$.
Moreover, sort $F_{l}$ by descending order with $|F_{1}|\geqslant|F_{2}|\geqslant\cdot\cdot\cdot\cdot
|F_{k}|\cdot\cdot\cdot\cdot\geqslant|F_{N}|$.
\begin{equation}
\label{eq:Ml} M_{l} = F_{l}[(x_{l}-x_{o})^2+(y_{l}-y_{o})^2],
\end{equation}
where ($x_{o},~y_{o}$) and ($x_{l},~y_{l}$) represent the galaxy's center and each pixel
position in Cartesian coordinates, respectively. Elliptical galaxies and galaxies
with bright nuclei have higher $Gini$ coefficient and lower $M_{20}$, while discs
and galaxies with a uniform surface brightness will have lower
$Gini$ coefficient and higher $M_{20}$.

The mean ellipticity and position of peak flux of the galaxy is measured using SExtractor \citep{Bertin1996}.
The non-parametric measurements and signal-to-noise estimations were performed counting the flux of pixels belonging
to a segmentation map. We defined the segmentation map by adopting the technique of \cite{Lotz2004},
where the pixels with surface brightness larger than the value at the Petrosian radii ($r_{p}$) measured in the smoothed image.
The elliptical $r_{p}$ corresponds to the semi-major axis where $I(r_{p})/\bar{I}(r<r_{p})=0.2$ \citep{Lotz2004}.
We calculated $M_{20}$ and $Gini$ coefficient by considering the pixels within these segmentation maps.

For galaxies at $1<z<3$, the above method has been used and tested by many previous
works \citep{Lotz2004,Lotz2006,Abraham2007,Fang2009,Fang2012,Fang2014,Kong2009,Wang2012,An2014}.
Benefiting from the stellar mass cut ($M_{*}\geq10^{10}M_{\odot}$), the median magnitude in the F160W band is $H_{\rm AB}\sim 22.9$.
And all their images have a mean S/N per pixel $\langle \rm S/N\rangle > 2$ with a median
of $\langle \rm S/N\rangle \sim 9$. Therefore, the $Gini$ and $M_{20}$ of galaxy for our sample
do not suffer S/N effect \citep{Lotz2004,Lee2013}.

\subsubsection{Non-parametric Morphology}

Figure 4 shows the results of the distribution on nonparametric morphology for eSFGs (blue), cSFGs (green) and cQGs (red) with $2\leq z\leq3$ and $M_{*}>10^{10}M_{\odot}$, cSFGs from GOODS-S field are also plotted in cyan squares. The cQGs and cSFGs have larger $Gini$ and smaller $M_{20}$, while eSFGs are reversed. We derived the mean ($M_{20}$, $Gini$) values for cQGs and cSFGs are ($-1.54\pm0.21$, $0.57\pm0.10$)
and ($-1.53\pm0.23$, $0.54\pm0.12$) respectively, whereas eSFGs are ($-1.38\pm0.29$, $0.43\pm0.09$). For the cSFGs from GOODS-S field,
the average values of $M_{20}$ and $Gini$ correspond to ($-1.55\pm0.18$, $0.57\pm0.08$).
We find the cSFGs distribute at the same locus as the cQGs on the $Gini$ vs. $M_{20}$ panel, but are obviously different from eSFGs. The similar distribution of morphology between cSFG and cQG indicates that there are
similar formation process for these galaxies. And from cSFG to cQG, the undisturbed quenching process such as gas-consuming, AGN or supernova feed back are the dominant mechanism to calm down the star formation activity in a compact phase.

\begin{figure}
\center
\includegraphics[width=1\columnwidth]{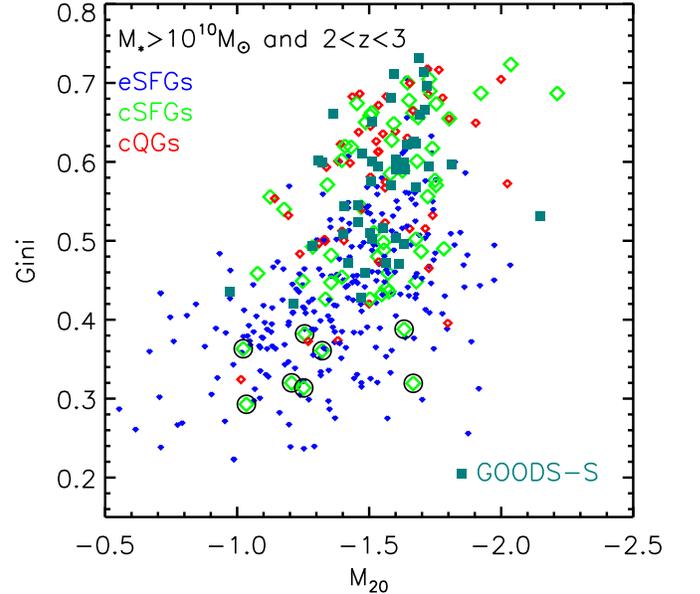}
\caption{Nonparametric morphology $Gini$ vs. $M_{20}$ for eSFGs (blue), cSFGs (green) and cQGs (red) with $2\leq z\leq3$ and $M_{*}>10^{10}M_{\odot}$. cSFGs from GOODS-S field are plotted in cyan squares. The cSFGs having $Gini<0.4$ are marked by black circles.}
\end{figure}

We present $HST$/WFC3 F160W images for cSFGs in Figure 5. The size of each postage map is $3.6''\times3.6''$, and $1''$ corresponds to $\sim8.5$ kpc at $z\sim2$. The source IDs, $Gini$ and $M_{20}$ are also labeled in each panel. The appearant morphology of cSFG in COSMOS field has really amazed us, we find about two thirds of cSFGs have spheroid morphologies (e.g., ID$=9,22,28$) the same as that of cQGs, and one third of them obviously show signatures of violent interactions or post mergers (e.g., ID$=16,34,36,37$). A small proportion of cSFGs are not visually ``compact'' and have $Gini<0.4$, the ID numbers of them are 18, 31, 34, 36, 48, 53, 56 and 59, as labeled in orange color in Figure 5. The consistency can be attributed to two reasons: on the one hand, some visually extended galaxies do have small effective radii but with small $Gini$, because the part outside the effective radius also contributes much to the measurement of $Gini$; on the other hand, the structural measurements from {\tt GALFIT} are unlikely to be accurate if galaxy has clump or multiple bright cores, and this factor could be used to clean out non-compact galaxies in the future larger samples. For example, it's very clear in Figure 5 that those visually extended cSFGs all have lower $Gini$ coefficients ($Gini<0.4$).

\begin{figure*}
\center
\includegraphics[width=0.95\textwidth]{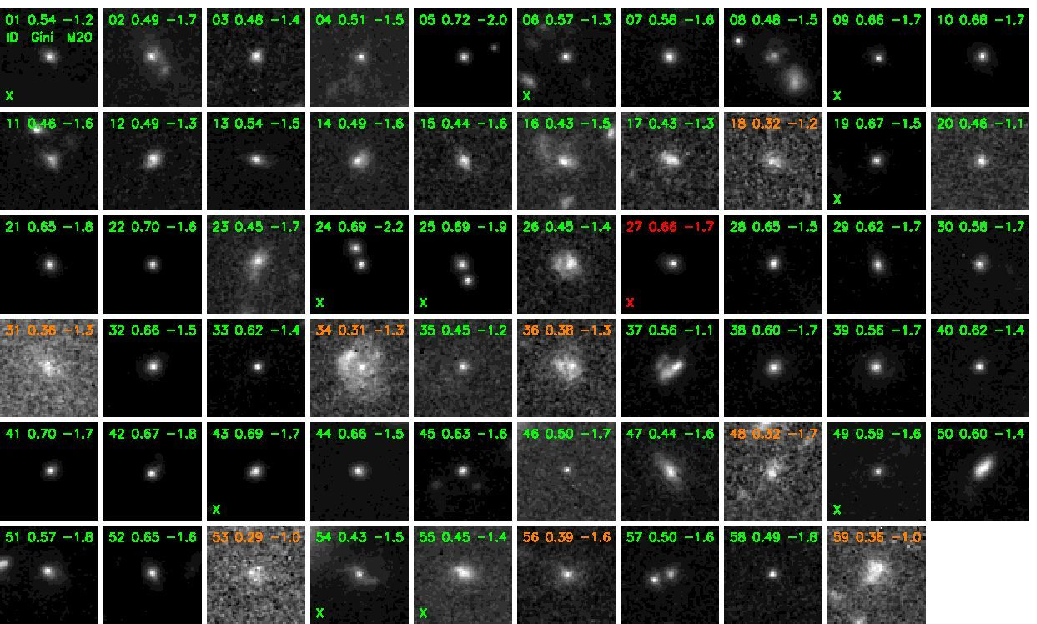}
\caption{$HST$/WFC3 F160W images for the 59 compact star-forming galaxies (cSFGs) in our sample selected from COSMOS field. The size of each postage map is $3.6''\times3.6''$, and $1''$ corresponds to $\sim8.5$ kpc at $z\sim2$. The source IDs, $Gini$ and $M_{20}$ are labeled for each galaxy in green color if it has $Gini\geq0.4$, otherwise we label them in orange, indicating they are visually extended with $Gini<0.4$. The cSFG labeled in red has been studied by spectroscopic measurements in Barro et al. (2014b). Sources having X-ray detections ($L_{0.5-10 \rm keV}>10^{41}$ erg s$^{-1}$) are marked with an ``X'' in the bottom-left corner.}
\end{figure*}

We also find nearly none of them can be recognized as disks. The large number of post-merger morphologies existing in cSFG sample implies that most of the progenitors of cSFGs have experienced violent gas-rich interactions such as dissipative wet mergers, which is considered to be one of the dominant mechanism for size shrink which makes masses distribute at a smaller radius and finally turns the eSFGs into cSFGs. While the disk instability is another dominant mechanism for shrinking galaxies which is simulated in \cite{Barro2014a}, but is not seen in our sample. This may caused by some of extended disk galaxies have already shrinked into these spheroids through disk instability at higher redshifts. Larger samples at higher redshifts are needed to exam whether the disk instability is an usual mechanism to shrink galaxies from eSFG to cSFG in the future work.

\subsection{Comparison among eSFG, cSFG, and cQG}

In Figure 6 we show the distributions on a series of physical parameters for eSFGs (blue), cSFGs (green) and cQGs (red) within $2\leq z\leq3$ and $M_{*}\geq 10^{10}M_{\odot}$. In this figure, the first three panels on the upper row are the best-fit stellar population parameters and the last three panels in the following line are the structural parameters. The median values for eSFGs, cSFGs (COSMOS+GOODS-S) and cQGs are marked by square points in corresponding colors. From these distributions, we find the difference of physical properties among eSFGs, cSFGs
and cQGs.

\begin{figure*}
\center
\includegraphics[width=0.99\textwidth]{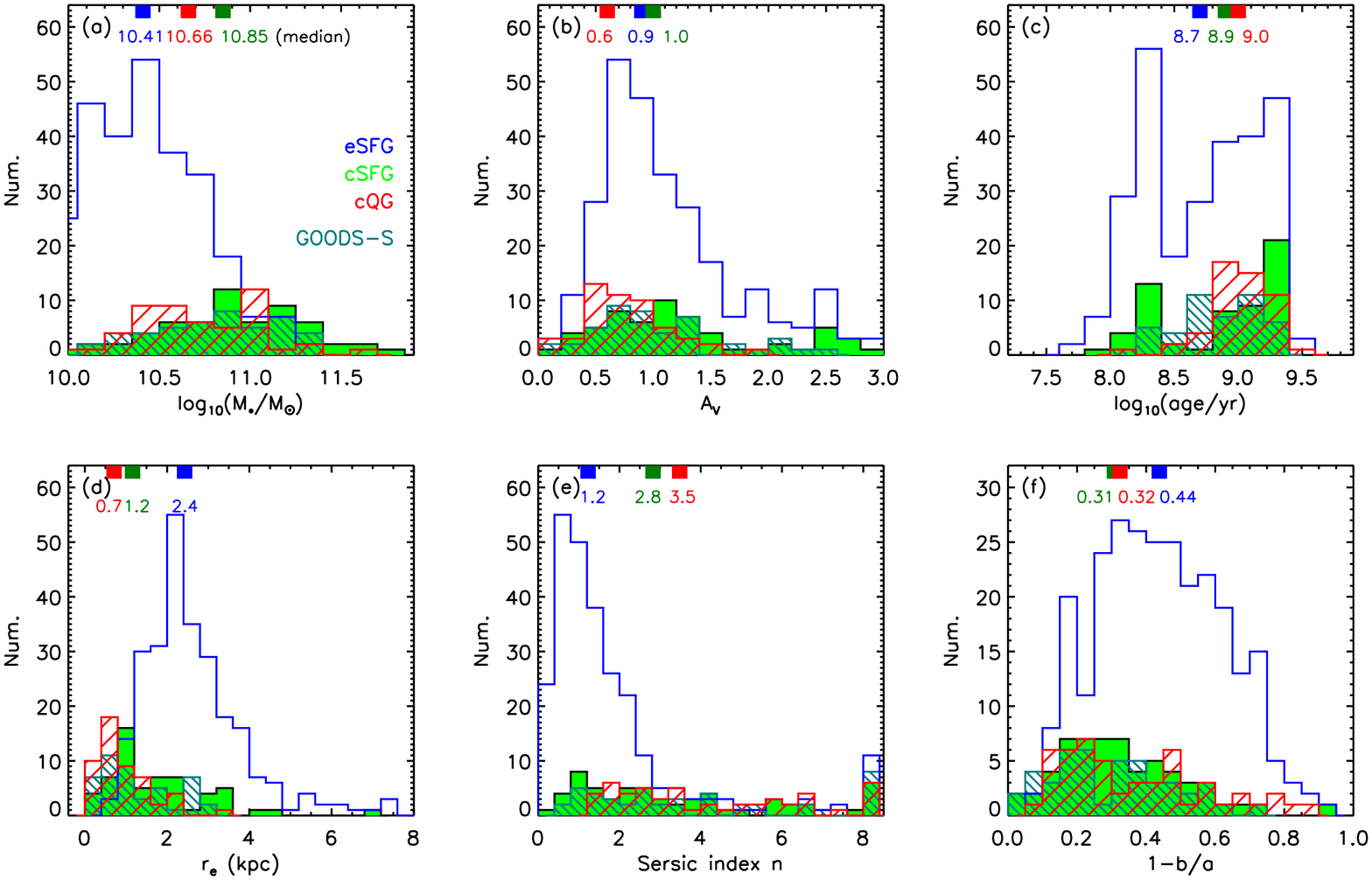}
\caption{Distributions on different physical parameters for eSFGs (blue), cSFGs (green) and cQGs (red) within $2\leq z\leq3$ and $M_{*}\geq 10^{10}M_{\odot}$. The first three panels on the upper row of this figure are the best-fit stellar population parameters: (a) stellar mass $M_{*}$, (b) extinction $A_{\mathrm V}$ and (c) stellar population age of galaxy. The last three panels in the following line are the structural parameters: (d) the effective radius $r_{\rm e}$, (e) the S\'{e}rsic index $n$ and (f) the ellipticity $1-b/a$. The 45 cSFGs selected from GOODS-S field are plotted in cyan color with oblique lines. The median values for eSFGs, cSFGs (COSMOS+GOODS-S) and cQGs are marked in blue, dark green and red square points on the top in each panel, respectively.}
\end{figure*}

The stellar masses of cSFGs and cQGs peaked at $\sim10^{10.5-11.5}M_{\odot}$, as shown by panel (a), they dominate the high mass end compared to eSFGs. This is a possible result if eSFGs merged into cSFGs, and then they quenched into cQGs. Considering cQGs come from higher redshifts and the global mass assembly of galaxies with the cosmic time, it is understandable that the median mass of cQGs are less massive than cSFGs in our redshift slice. Extinctions correlate with SFRs of galaxies \citep{Reddy2010}.
Panel (b) shows that cSFGs have heaviest and cQGs have fewest extinctions, and eSFGs are in the middle. From this result we infer the evolutionary scenario that no matter through merger or disk instability the eSFGs evolved into cSFGs, during this period the star formation activity becomes fierce, and the extinctions also increase with it. When cSFGs quenched into cQGs, most of gas has been used up, the extinction of galaxy thus decreased. Panel (c) shows the stellar age of eSFGs, cSFGs and cQGs, and not surprisingly, cSFGs and cQGs have the oldest ages. The median age of eSFGs is $10^{8.7}$ yr, and the median ages of cSFGs and cQGs are $10^{8.9}$ and $10^{9.0}$ yr, respectively.

Panel (d), (e) and (f) show the structural parameters for eSFGs, cSFGs and cQGs. The results of these distributions are all understandable if we assume that cSFG is a transitional type of galaxy between eSFG and cQG. Through merger or disk instability, eSFG turns into cSFG, the masses redistribute at a smaller radius, thus from eSFG, cSFG to cQG, the $r_{\rm e}$ becomes smaller and the S\'{e}rsic index $n$ becomes larger. The distributions on ellipticity $1-b/a$ indicate that the morphology of most of eSFGs have been disturbed by external interactions such as wet mergers, which also does not support the disk instability mechanism is common.

Based on the analysis above, we find the general distributions of cSFGs on different physical parameters are very similar to that of cQGs, as shown by Figure 6. The existence of the similarity on physical properties between cSFGs and cQGs within $2\leq z\leq3$ and $M_{*}\geq 10^{10}M_{\odot}$ is consistent with our speculation
that the cSFGs are short-lived and will rapidly quenched into cQGs. In Table 1, we list all the main physical parameters previously mentioned for 59 compact SFGs in the COSMOS field. The median values of physical parameters of cSFGs in
our sample are as follows: $\langle{\rm log} (M_{\ast}/M_{\odot})\rangle=10.89\pm0.39$, $\langle A_{\rm V}/{\rm mag}\rangle=1.1\pm0.8$,
$\langle{\rm log} (\rm age/{\rm yr}) \rangle=9.0\pm0.5$, $\langle r_{\rm e}/{\rm kpc}\rangle=1.3\pm1.2$,
$\langle n\rangle=2.8\pm2.4$, and $\langle 1-b/a\rangle=0.30\pm0.18$, in good agreement with those provided by \cite{Barro2014a}
($\langle{\rm log} (M_{\ast}/M_{\odot})\rangle=10.82\pm0.33$, $\langle A_{\rm V}/{\rm mag}\rangle=0.9\pm0.6$,
$\langle{\rm log} (\rm age/{\rm yr}) \rangle=8.8\pm0.3$, $\langle r_{\rm e}/{\rm kpc}\rangle=1.0\pm0.9$,
$\langle n\rangle=3.1\pm2.5$, and $\langle 1-b/a\rangle=0.31\pm0.18$).

\begin{table*}
\scriptsize
\setlength{\tabcolsep}{0.011in}
\begin{minipage}{181mm}
\caption{Physical parameters for 59 compact SFGs in the COSMOS field.}
\begin{tabular}{@{}lcccccccccccccccc@{}}
\hline
  ID   &  R.A. & Dec. & $z(\Delta z^{c})$ & $\rm log~SFR$ & $\rm log~$$M_{*}$ & $A_{\mathrm V}$ & $\rm log~age$ & $r_{\rm e}$ & $n$ & $b/a$ & $Gini$ & $M_{20}$ & $U-V$ & $V-J$ & $U-B$ & $M_{\rm B}$   \\
    & (deg.) & (deg.) & & ($M_{\odot}$ $\rm yr^{-1}$) & ($M_{\odot}$) & (mag) & (yr) & (kpc) & & & & & (mag) & (mag) & (mag) & (mag) \\
  (1)  &  (2)  & (3)  &  (4)  &  (5)  &  (6) &  (7) &  (8) &  (9) &  (10) &  (11) &  (12) &  (13) &  (14)  &  (15)  &  (16)  &  (17)   \\
\hline
01& 150.104218& 2.178912& 2.32(-0.19)& 2.61$\pm$0.03& 10.85$\pm$0.23& 1.1$\pm$0.2& 9.2$\pm$0.3& 1.11$\pm$0.05& 2.52$\pm$0.35& 0.60$\pm$0.04&  0.54$\pm$0.01& -1.18$\pm$0.08&  1.22$\pm$0.12&  1.60$\pm$0.10&  0.77$\pm$0.06& -21.68$\pm$0.06 \\
02& 150.103958& 2.186443& 2.55(~0.20)& 2.48$\pm$0.04& 11.61$\pm$0.05& 0.9$\pm$0.3& 9.4$\pm$0.3& 2.81$\pm$0.06& 2.79$\pm$0.25& 0.72$\pm$0.02&  0.49$\pm$0.02& -1.70$\pm$0.08&  1.78$\pm$0.21&  1.77$\pm$0.20&  1.08$\pm$0.09& -22.65$\pm$0.05 \\
03& 150.081833& 2.186991& 2.11(-0.30)& 2.49$\pm$0.03& 10.46$\pm$0.12& 2.3$\pm$0.3& 8.4$\pm$0.2& 0.88$\pm$0.05& 1.57$\pm$0.35& 0.65$\pm$0.06&  0.48$\pm$0.01& -1.36$\pm$0.10&  1.54$\pm$0.10&  1.84$\pm$0.15&  1.00$\pm$0.12& -20.22$\pm$0.06 \\
04& 150.119690& 2.188866& 2.30(-0.06)& 1.05$\pm$0.13& 10.54$\pm$0.10& 1.3$\pm$0.1& 8.9$\pm$0.4& 1.04$\pm$0.08& 6.11$\pm$1.99& 0.81$\pm$0.10&  0.51$\pm$0.01& -1.52$\pm$0.08&  1.76$\pm$0.13&  1.17$\pm$0.09&  1.08$\pm$0.10& -20.89$\pm$0.09 \\
05& 150.149673& 2.191054& 2.15(~0.04)& 1.45$\pm$0.12& 10.71$\pm$0.26& 1.1$\pm$0.2& 8.0$\pm$0.1& 0.74$\pm$0.01& 8.00$\pm$0.34& 0.79$\pm$0.02&  0.72$\pm$0.03& -2.04$\pm$0.10&  0.75$\pm$0.11&  0.77$\pm$0.12&  0.55$\pm$0.16& -23.26$\pm$0.03 \\
06& 150.193451& 2.199375& 2.18(~0.01)& 1.84$\pm$0.08& 10.50$\pm$0.08& 0.1$\pm$0.1& 9.4$\pm$0.5& 0.60$\pm$0.04& 6.01$\pm$1.29& 0.70$\pm$0.07&  0.57$\pm$0.01& -1.34$\pm$0.05&  1.53$\pm$0.15&  0.94$\pm$0.15&  0.86$\pm$0.12& -21.09$\pm$0.14 \\
07& 150.061646& 2.205388& 2.26(~0.22)& 1.20$\pm$0.19& 10.87$\pm$0.14& 0.7$\pm$0.3& 9.2$\pm$0.3& 1.22$\pm$0.02& 2.04$\pm$0.14& 0.97$\pm$0.02&  0.58$\pm$0.01& -1.58$\pm$0.07&  1.50$\pm$0.23&  1.17$\pm$0.06&  0.89$\pm$0.08& -21.84$\pm$0.05 \\
08& 150.076004& 2.211829& 2.20(~0.03)& 2.23$\pm$0.04& 11.36$\pm$0.11& 2.7$\pm$0.4& 8.3$\pm$0.2& 3.55$\pm$0.04& 2.01$\pm$0.09& 0.69$\pm$0.01&  0.48$\pm$0.02& -1.53$\pm$0.09&  1.66$\pm$0.09&  1.93$\pm$0.23&  1.01$\pm$0.12& -22.21$\pm$0.04 \\
09& 150.192642& 2.219852& 2.87(-0.04)& 2.41$\pm$0.03& 10.83$\pm$0.11& 0.2$\pm$0.1& 9.3$\pm$0.1& 0.96$\pm$0.02& 8.00$\pm$0.76& 0.83$\pm$0.03&  0.66$\pm$0.03& -1.67$\pm$0.05&  0.70$\pm$0.20&  1.24$\pm$0.08&  0.39$\pm$0.16& -23.12$\pm$0.05 \\
10& 150.117264& 2.223891& 2.06(~0.09)& 1.86$\pm$0.06& 10.93$\pm$0.06& 1.0$\pm$0.2& 9.0$\pm$0.3& 1.11$\pm$0.02& 4.46$\pm$0.24& 0.88$\pm$0.02&  0.68$\pm$0.02& -1.65$\pm$0.11&  1.20$\pm$0.08&  1.06$\pm$0.11&  0.79$\pm$0.07& -22.51$\pm$0.07 \\
11& 150.073120& 2.233184& 2.83(~0.03)& 2.43$\pm$0.03& 10.74$\pm$0.10& 0.6$\pm$0.2& 9.2$\pm$0.2& 1.50$\pm$0.13& 0.42$\pm$0.21& 0.81$\pm$0.11&  0.46$\pm$0.01& -1.57$\pm$0.07&  1.77$\pm$0.27&  1.20$\pm$0.05&  1.06$\pm$0.13& -21.32$\pm$0.02 \\
12& 150.179230& 2.233675& 2.17(-0.44)& 2.51$\pm$0.02& 10.84$\pm$0.05& 1.8$\pm$0.3& 8.8$\pm$0.4& 1.68$\pm$0.05& 0.99$\pm$0.10& 0.51$\pm$0.03&  0.49$\pm$0.01& -1.29$\pm$0.04&  1.85$\pm$0.14&  1.59$\pm$0.13&  1.12$\pm$0.09& -21.10$\pm$0.08 \\
13& 150.154510& 2.233613& 2.24(~0.24)& 1.62$\pm$0.11& 11.00$\pm$0.08& 2.4$\pm$0.3& 8.4$\pm$0.2& 1.34$\pm$0.08& 4.21$\pm$0.49& 0.27$\pm$0.03&  0.54$\pm$0.01& -1.47$\pm$0.06&  1.74$\pm$0.06&  1.89$\pm$0.22&  1.06$\pm$0.05& -21.41$\pm$0.06 \\
14& 150.171860& 2.240702& 2.35(-0.09)& 2.44$\pm$0.02& 11.16$\pm$0.03& 2.6$\pm$0.4& 8.3$\pm$0.1& 2.34$\pm$0.05& 1.50$\pm$0.14& 0.83$\pm$0.02&  0.49$\pm$0.02& -1.56$\pm$0.14&  1.61$\pm$0.11&  1.91$\pm$0.09&  0.99$\pm$0.11& -21.82$\pm$0.05 \\
15& 150.198013& 2.244380& 2.03(~0.17)& 1.36$\pm$0.17& 10.71$\pm$0.22& 1.6$\pm$0.2& 9.4$\pm$0.3& 1.42$\pm$0.09& 0.99$\pm$0.24& 0.43$\pm$0.05&  0.44$\pm$0.01& -1.56$\pm$0.09&  1.77$\pm$0.13&  1.79$\pm$0.16&  1.02$\pm$0.07& -20.31$\pm$0.06 \\
16& 150.106338& 2.251579& 2.73(~0.01)& 2.64$\pm$0.03& 11.36$\pm$0.05& 2.9$\pm$0.4& 8.0$\pm$0.2& 4.32$\pm$0.15& 3.00$\pm$0.46& 0.79$\pm$0.03&  0.43$\pm$0.01& -1.55$\pm$0.11&  1.73$\pm$0.10&  1.95$\pm$0.07&  1.03$\pm$0.14& -22.49$\pm$0.10 \\
17& 150.130005& 2.252692& 2.82(~0.33)& 2.37$\pm$0.02& 11.06$\pm$0.07& 0.9$\pm$0.1& 9.3$\pm$0.2& 1.79$\pm$0.08& 0.85$\pm$0.22& 0.65$\pm$0.05&  0.43$\pm$0.01& -1.33$\pm$0.06&  1.80$\pm$0.21&  1.56$\pm$0.09&  1.09$\pm$0.15& -21.68$\pm$0.03 \\
$18^a$& 150.149231& 2.254018& 2.73(~0.16)& 1.67$\pm$0.09& 10.85$\pm$0.04& 0.7$\pm$0.2& 9.3$\pm$0.1& 1.90$\pm$0.20& 0.98$\pm$0.44& 0.53$\pm$0.09&  0.32$\pm$0.03& -1.21$\pm$0.05&  1.95$\pm$0.16&  1.55$\pm$0.13&  1.18$\pm$0.21& -21.00$\pm$0.16 \\
19& 150.141998& 2.265100& 2.06(-0.36)& 2.28$\pm$0.03& 10.84$\pm$0.12& 0.5$\pm$0.1& 9.4$\pm$0.5& 0.64$\pm$0.02& 4.16$\pm$0.43& 0.88$\pm$0.03&  0.67$\pm$0.02& -1.45$\pm$0.14&  1.71$\pm$0.08&  1.40$\pm$0.06&  0.98$\pm$0.09& -21.43$\pm$0.04 \\
20& 150.085083& 2.272684& 2.22(~0.02)& 1.12$\pm$0.14& 10.47$\pm$0.07& 1.5$\pm$0.3& 9.3$\pm$0.3& 0.94$\pm$0.08& 1.72$\pm$0.66& 0.74$\pm$0.10&  0.46$\pm$0.01& -1.08$\pm$0.02&  1.30$\pm$0.12&  1.68$\pm$0.13&  0.75$\pm$0.06& -20.15$\pm$0.07 \\
21& 150.142944& 2.278515& 2.24(~0.02)& 1.58$\pm$0.10& 10.88$\pm$0.09& 0.4$\pm$0.2& 9.0$\pm$0.1& 1.70$\pm$0.03& 5.78$\pm$0.40& 0.89$\pm$0.02&  0.65$\pm$0.01& -1.80$\pm$0.15&  1.66$\pm$0.10&  0.98$\pm$0.18&  1.09$\pm$0.16& -22.33$\pm$0.02 \\
22& 150.164093& 2.294364& 2.12(-0.04)& 0.78$\pm$0.35& 10.51$\pm$0.11& 0.3$\pm$0.1& 9.0$\pm$0.3& 0.45$\pm$0.01& 5.75$\pm$0.51& 0.88$\pm$0.03&  0.70$\pm$0.03& -1.65$\pm$0.12&  1.20$\pm$0.09&  0.73$\pm$0.12&  0.82$\pm$0.12& -21.97$\pm$0.08 \\
23& 150.155319& 2.295565& 2.34(-0.11)& 2.22$\pm$0.04& 11.27$\pm$0.13& 2.6$\pm$0.2& 8.3$\pm$0.2& 3.25$\pm$0.10& 2.08$\pm$0.25& 0.59$\pm$0.02&  0.45$\pm$0.01& -1.68$\pm$0.08&  1.54$\pm$0.11&  2.08$\pm$0.25&  0.98$\pm$0.06& -22.01$\pm$0.05 \\
24& 150.119568& 2.295787& 2.19(~0.11)& 2.46$\pm$0.02& 11.23$\pm$0.09& 1.1$\pm$0.3& 8.9$\pm$0.5& 1.00$\pm$0.02& 8.00$\pm$0.39& 0.61$\pm$0.02&  0.69$\pm$0.02& -2.21$\pm$0.23&  1.30$\pm$0.07&  0.82$\pm$0.08&  0.79$\pm$0.10& -23.10$\pm$0.09 \\
25& 150.119614& 2.295948& 2.12(~0.04)& 2.46$\pm$0.03& 11.07$\pm$0.07& 1.1$\pm$0.1& 8.5$\pm$0.2& 2.10$\pm$0.02& 5.16$\pm$0.18& 0.76$\pm$0.01&  0.69$\pm$0.01& -1.92$\pm$0.13&  1.22$\pm$0.05&  0.80$\pm$0.10&  0.83$\pm$0.08& -23.28$\pm$0.03 \\
26& 150.167709& 2.298764& 2.80(-0.01)& 2.56$\pm$0.03& 11.50$\pm$0.09& 0.8$\pm$0.2& 9.3$\pm$0.3& 2.81$\pm$0.04& 0.49$\pm$0.07& 0.82$\pm$0.02&  0.45$\pm$0.01& -1.36$\pm$0.07&  1.82$\pm$0.13&  1.74$\pm$0.12&  1.12$\pm$0.11& -22.64$\pm$0.12 \\
$27^b$& 150.074615& 2.302011& 2.15(-0.02)& 2.22$\pm$0.04& 11.22$\pm$0.12& 1.0$\pm$0.3& 9.1$\pm$0.1& 2.07$\pm$0.04& 5.68$\pm$0.30& 0.57$\pm$0.02&  0.66$\pm$0.03& -1.69$\pm$0.03&  1.67$\pm$0.20&  1.40$\pm$0.07&  1.04$\pm$0.09& -22.49$\pm$0.06 \\
28& 150.118744& 2.302693& 2.61(~0.10)& 1.89$\pm$0.07& 10.23$\pm$0.08& 0.6$\pm$0.2& 8.4$\pm$0.2& 0.73$\pm$0.02& 1.75$\pm$0.18& 0.65$\pm$0.03&  0.65$\pm$0.02& -1.49$\pm$0.06&  0.76$\pm$0.10&  0.27$\pm$0.12&  0.49$\pm$0.17& -22.29$\pm$0.03 \\
29& 150.177582& 2.305114& 2.19(-0.10)& 2.06$\pm$0.04& 10.79$\pm$0.18& 0.6$\pm$0.1& 9.2$\pm$0.6& 0.97$\pm$0.03& 2.78$\pm$0.21& 0.43$\pm$0.02&  0.62$\pm$0.02& -1.74$\pm$0.10&  1.49$\pm$0.11&  1.05$\pm$0.06&  0.91$\pm$0.06& -21.80$\pm$0.07 \\
30& 150.166122& 2.307469& 2.75(-0.06)& 1.75$\pm$0.09& 10.74$\pm$0.06& 1.4$\pm$0.4& 8.8$\pm$0.3& 0.91$\pm$0.03& 2.40$\pm$0.28& 0.75$\pm$0.03&  0.58$\pm$0.01& -1.75$\pm$0.11&  1.21$\pm$0.12&  1.16$\pm$0.11&  0.77$\pm$0.12& -22.12$\pm$0.02 \\
$31^a$& 150.098648& 2.311156& 2.94(~0.00)& 2.82$\pm$0.04& 11.29$\pm$0.09& 1.5$\pm$0.2& 9.2$\pm$0.4& 3.33$\pm$0.25& 1.18$\pm$0.40& 0.55$\pm$0.06&  0.36$\pm$0.03& -1.32$\pm$0.02&  1.69$\pm$0.15&  1.84$\pm$0.09&  1.06$\pm$0.07& -22.03$\pm$0.11 \\
32& 150.099686& 2.311813& 2.45(-0.04)& 2.07$\pm$0.05& 11.00$\pm$0.07& 1.0$\pm$0.1& 8.7$\pm$0.2& 1.23$\pm$0.02& 2.47$\pm$0.14& 0.77$\pm$0.02&  0.66$\pm$0.01& -1.51$\pm$0.05&  1.50$\pm$0.08&  0.93$\pm$0.13&  0.98$\pm$0.13& -22.66$\pm$0.04 \\
33& 150.068192& 2.334341& 2.52(~0.07)& 1.79$\pm$0.06& 10.45$\pm$0.31& 0.4$\pm$0.2& 8.9$\pm$0.1& 0.25$\pm$0.01& 7.87$\pm$1.13& 0.76$\pm$0.05&  0.62$\pm$0.01& -1.43$\pm$0.08&  1.32$\pm$0.07&  0.69$\pm$0.16&  0.89$\pm$0.07& -21.77$\pm$0.05 \\
$34^a$& 150.122528& 2.347118& 2.43(~0.01)& 1.91$\pm$0.05& 10.75$\pm$0.09& 1.1$\pm$0.1& 8.5$\pm$0.3& 0.94$\pm$0.03& 8.00$\pm$1.30& 0.99$\pm$0.06&  0.31$\pm$0.04& -1.25$\pm$0.02&  1.05$\pm$0.10&  0.78$\pm$0.11&  0.68$\pm$0.14& -22.70$\pm$0.14 \\
35& 150.173660& 2.358639& 2.98(-0.08)& 2.10$\pm$0.04& 10.40$\pm$0.05& 0.6$\pm$0.1& 8.8$\pm$0.2& 0.84$\pm$0.08& 4.12$\pm$1.48& 0.75$\pm$0.11&  0.45$\pm$0.01& -1.25$\pm$0.07&  1.25$\pm$0.14&  0.52$\pm$0.20&  0.81$\pm$0.08& -21.74$\pm$0.03 \\
$36^a$& 150.098541& 2.365358& 2.82(~0.00)& 2.83$\pm$0.03& 11.34$\pm$0.06& 2.6$\pm$0.2& 8.2$\pm$0.1& 3.39$\pm$0.07& 0.60$\pm$0.11& 0.82$\pm$0.03&  0.38$\pm$0.03& -1.26$\pm$0.09&  1.65$\pm$0.15&  1.85$\pm$0.06&  1.04$\pm$0.07& -22.48$\pm$0.06 \\
37& 150.163559& 2.372434& 2.03(~0.00)& 2.63$\pm$0.03& 10.89$\pm$0.06& 1.1$\pm$0.2& 9.2$\pm$0.2& 1.87$\pm$0.03& 0.20$\pm$0.01& 0.30$\pm$0.01&  0.56$\pm$0.01& -1.13$\pm$0.03&  0.96$\pm$0.16&  1.46$\pm$0.12&  0.63$\pm$0.15& -22.51$\pm$0.08 \\
38& 150.111176& 2.373299& 2.45(-0.05)& 1.48$\pm$0.10& 11.15$\pm$0.11& 0.6$\pm$0.1& 9.2$\pm$0.3& 1.73$\pm$0.03& 2.39$\pm$0.15& 0.74$\pm$0.02&  0.60$\pm$0.01& -1.68$\pm$0.09&  1.88$\pm$0.18&  1.12$\pm$0.08&  1.17$\pm$0.11& -22.44$\pm$0.05 \\
39& 150.187180& 2.380155& 2.54(~0.00)& 2.41$\pm$0.05& 11.19$\pm$0.09& 0.7$\pm$0.2& 9.1$\pm$0.4& 2.53$\pm$0.04& 3.68$\pm$0.26& 0.93$\pm$0.02&  0.56$\pm$0.01& -1.72$\pm$0.12&  1.54$\pm$0.06&  1.20$\pm$0.10&  0.99$\pm$0.17& -22.80$\pm$0.12 \\
40& 150.163818& 2.381426& 2.56(~0.04)& 1.00$\pm$0.22& 10.52$\pm$0.14& 0.3$\pm$0.2& 9.1$\pm$0.1& 0.80$\pm$0.04& 8.00$\pm$1.51& 0.70$\pm$0.06&  0.62$\pm$0.02& -1.41$\pm$0.03&  1.50$\pm$0.13&  0.76$\pm$0.07&  0.98$\pm$0.20& -21.67$\pm$0.04 \\
41& 150.060349& 2.382773& 2.47(~0.09)& 2.23$\pm$0.04& 10.53$\pm$0.22& 0.5$\pm$0.1& 8.1$\pm$0.3& 1.05$\pm$0.02& 6.66$\pm$0.32& 0.63$\pm$0.02&  0.70$\pm$0.01& -1.73$\pm$0.05&  0.70$\pm$0.10&  0.22$\pm$0.11&  0.49$\pm$0.17& -23.29$\pm$0.07 \\
42& 150.068481& 2.383552& 2.31(-0.09)& 2.42$\pm$0.04& 10.58$\pm$0.17& 0.8$\pm$0.3& 8.4$\pm$0.3& 1.11$\pm$0.02& 6.41$\pm$0.36& 0.58$\pm$0.02&  0.67$\pm$0.01& -1.75$\pm$0.07&  0.74$\pm$0.14&  0.59$\pm$0.09&  0.50$\pm$0.19& -22.84$\pm$0.05 \\
43& 150.071472& 2.414533& 2.47(-0.01)& 2.23$\pm$0.03& 11.10$\pm$0.06& 0.9$\pm$0.2& 8.9$\pm$0.2& 1.06$\pm$0.02& 2.82$\pm$0.15& 0.67$\pm$0.02&  0.69$\pm$0.02& -1.73$\pm$0.06&  1.24$\pm$0.07&  0.88$\pm$0.13&  0.78$\pm$0.09& -23.01$\pm$0.02 \\
44& 150.068146& 2.415469& 2.52(~0.02)& 1.58$\pm$0.07& 10.85$\pm$0.15& 0.3$\pm$0.1& 9.0$\pm$0.3& 0.95$\pm$0.02& 2.51$\pm$0.19& 0.61$\pm$0.02&  0.66$\pm$0.02& -1.50$\pm$0.10&  1.61$\pm$0.18&  0.77$\pm$0.06&  1.06$\pm$0.07& -22.43$\pm$0.08 \\
45& 150.105194& 2.416420& 2.52(-0.02)& 1.71$\pm$0.06& 10.20$\pm$0.04& 0.9$\pm$0.2& 8.2$\pm$0.1& 0.54$\pm$0.02& 1.81$\pm$0.27& 0.70$\pm$0.04&  0.63$\pm$0.01& -1.59$\pm$0.07&  0.79$\pm$0.09&  0.62$\pm$0.14&  0.57$\pm$0.12& -21.91$\pm$0.15 \\
46& 150.116623& 2.440312& 2.49(-0.10)& 0.30$\pm$0.22& 10.01$\pm$0.13& 0.4$\pm$0.1& 9.4$\pm$0.4& 0.04$\pm$0.04& 3.24$\pm$1.32& 0.06$\pm$0.14&  0.50$\pm$0.01& -1.68$\pm$0.12&  1.94$\pm$0.14&  0.59$\pm$0.15&  1.13$\pm$0.06& -20.01$\pm$0.05 \\
47& 150.111389& 2.452996& 2.28(~0.20)& 2.12$\pm$0.04& 11.34$\pm$0.08& 1.2$\pm$0.3& 9.2$\pm$0.3& 3.06$\pm$0.10& 1.58$\pm$0.15& 0.45$\pm$0.02&  0.44$\pm$0.02& -1.57$\pm$0.03&  1.87$\pm$0.11&  1.89$\pm$0.05&  1.14$\pm$0.10& -22.12$\pm$0.07 \\
$48^a$& 150.150284& 2.454510& 2.78(~0.23)& 2.32$\pm$0.04& 11.17$\pm$0.07& 1.4$\pm$0.2& 9.2$\pm$0.5& 1.92$\pm$0.20& 0.98$\pm$0.41& 0.48$\pm$0.08&  0.32$\pm$0.04& -1.67$\pm$0.02&  1.79$\pm$0.08&  1.54$\pm$0.14&  1.10$\pm$0.15& -21.85$\pm$0.03 \\
49& 150.125122& 2.455784& 2.26(~0.15)& 1.39$\pm$0.13& 10.12$\pm$0.16& 1.2$\pm$0.1& 8.1$\pm$0.1& 0.36$\pm$0.02& 8.00$\pm$1.63& 0.62$\pm$0.07&  0.59$\pm$0.01& -1.63$\pm$0.13&  0.98$\pm$0.18&  0.83$\pm$0.08&  0.62$\pm$0.09& -21.40$\pm$0.06 \\
50& 150.189255& 2.460574& 2.07(~0.23)& 1.79$\pm$0.08& 11.22$\pm$0.09& 1.3$\pm$0.2& 8.9$\pm$0.3& 2.04$\pm$0.03& 1.20$\pm$0.05& 0.38$\pm$0.01&  0.60$\pm$0.02& -1.40$\pm$0.02&  1.65$\pm$0.05&  1.20$\pm$0.10&  1.00$\pm$0.11& -22.57$\pm$0.14 \\
51& 150.173233& 2.464311& 2.16(~0.25)& 2.35$\pm$0.03& 11.08$\pm$0.24& 2.1$\pm$0.3& 8.4$\pm$0.2& 2.31$\pm$0.04& 3.33$\pm$0.23& 0.81$\pm$0.02&  0.57$\pm$0.01& -1.75$\pm$0.11&  1.33$\pm$0.12&  1.53$\pm$0.12&  0.82$\pm$0.07& -22.54$\pm$0.06 \\
52& 150.080872& 2.470742& 2.09(~0.15)& 1.93$\pm$0.07& 10.96$\pm$0.11& 1.3$\pm$0.2& 9.0$\pm$0.3& 1.15$\pm$0.03& 2.93$\pm$0.17& 0.50$\pm$0.02&  0.65$\pm$0.01& -1.59$\pm$0.04&  1.23$\pm$0.06&  1.26$\pm$0.08&  0.77$\pm$0.16& -22.39$\pm$0.04 \\
$53^a$& 150.135178& 2.479366& 2.82(~0.08)& 2.16$\pm$0.04& 11.13$\pm$0.05& 1.1$\pm$0.2& 9.3$\pm$0.6& 2.93$\pm$0.21& 1.26$\pm$0.59& 0.92$\pm$0.09&  0.29$\pm$0.04& -1.03$\pm$0.02&  1.79$\pm$0.19&  1.64$\pm$0.15&  1.09$\pm$0.08& -21.65$\pm$0.07 \\
54& 150.059204& 2.505265& 2.94(-0.20)& 3.79$\pm$0.03& 11.73$\pm$0.07& 2.7$\pm$0.3& 7.9$\pm$0.1& 4.56$\pm$0.13& 3.68$\pm$0.32& 0.46$\pm$0.02&  0.43$\pm$0.01& -1.50$\pm$0.03&  1.48$\pm$0.15&  1.85$\pm$0.11&  0.91$\pm$0.13& -23.95$\pm$0.09 \\
55& 150.093384& 2.507325& 2.55(~0.12)& 2.54$\pm$0.02& 10.95$\pm$0.11& 2.1$\pm$0.4& 8.2$\pm$0.2& 2.22$\pm$0.06& 1.11$\pm$0.12& 0.54$\pm$0.02&  0.45$\pm$0.01& -1.40$\pm$0.02&  1.37$\pm$0.06&  1.64$\pm$0.07&  0.85$\pm$0.06& -22.33$\pm$0.16 \\
$56^a$& 150.137497& 2.513580& 2.89(~0.35)& 2.73$\pm$0.03& 11.70$\pm$0.23& 1.2$\pm$0.1& 9.2$\pm$0.4& 7.04$\pm$0.33& 3.91$\pm$0.76& 0.70$\pm$0.04&  0.39$\pm$0.02& -1.63$\pm$0.09&  1.90$\pm$0.20&  1.73$\pm$0.12&  1.16$\pm$0.12& -23.01$\pm$0.11 \\
57& 150.106964& 2.524392& 2.39(~0.10)& 2.04$\pm$0.06& 10.95$\pm$0.08& 2.4$\pm$0.2& 8.2$\pm$0.1& 2.05$\pm$0.09& 4.10$\pm$0.71& 0.76$\pm$0.04&  0.50$\pm$0.01& -1.55$\pm$0.05&  1.55$\pm$0.06&  1.83$\pm$0.09&  0.96$\pm$0.07& -21.67$\pm$0.06 \\
58& 150.142181& 2.531121& 2.18(-0.08)& 1.77$\pm$0.09& 10.34$\pm$0.05& 1.3$\pm$0.1& 9.0$\pm$0.2& 0.35$\pm$0.04& 1.02$\pm$0.39& 0.57$\pm$0.10&  0.49$\pm$0.02& -1.78$\pm$0.12&  1.52$\pm$0.12&  1.41$\pm$0.06&  0.94$\pm$0.13& -20.22$\pm$0.09 \\
$59^a$& 150.075516& 2.544108& 2.35(-0.07)& 2.13$\pm$0.05& 11.44$\pm$0.07& 2.7$\pm$0.4& 8.4$\pm$0.3& 3.30$\pm$0.08& 0.73$\pm$0.12& 0.67$\pm$0.03&  0.36$\pm$0.03& -1.02$\pm$0.02&  1.64$\pm$0.08&  2.21$\pm$0.16&  1.08$\pm$0.09& -22.27$\pm$0.03 \\
\hline
\end{tabular}

Notes: \\ $^a$ 8 cSFGs with $Gini<0.4$ and extended visual morphologies. \\ $^b$ cSFG with spectroscopic redshift $z_{\rm spec}=2.095$ in Barro et al. (2014b).\\ $^c$ $\Delta z=z_{\rm 3D-HST}-z_{\rm UltraVISTA}$.
\end{minipage}
\end{table*}

\subsection{Fraction of AGNs}

We expect the transformation from extended SFGs to compact SFGs will trigger both star formation and black hole growth, thus higher fraction of AGNs should be detected in cSFG sample. There have been several studies of AGN selection using mid-IR color or other IR properties, the mid-IR photometry has been proven to be a robust and efficient tool to select AGNs as their properties at these wavelengths are typically very different from those of stars and galaxies \citep{Stern2005,Donley2008,Park2010}. In this section, three different AGN selection methods are employed to identify AGNs in our sample.

Firstly, we select AGNs using the criterion defined by \cite{Stern2005}, which were based on the spectroscopic sample of the AGN and Galaxy Evolution Survey. Figure 7 shows the IRAC [5.8]-[8.0] vs. [3.6]-[4.5] color space for galaxies in the COSMOS-CANDELS field. The solid purple lines indicate the boundaries of the AGN selection region introduced by \cite{Stern2005}. Small points represent all galaxies with IRAC measurements in this field, eSFGs, cSFGs and cQGs are shown by blue, green and red squares, respectively. In our sample, 269 eSFGs, 59 cSFGs and 51 cQGs have been detected in all four IRAC channels, among them, 61 eSFGs, 16 cSFGs and 8 cQGs are identified as AGN candidates, the AGN fraction of them are 23\%, 27\% and 16\%, respectively. The fraction of AGN in cSFGs is slightly higher than that in eSFGs and cQGs, which supports the expectation that during the transformation from extended SFGs to compact SFGs (no matter by merging or disk instability), the inflow of large amount of gas will intensify the activity level of the black hole in galaxy center. The mid-IR color criterion is reliable for separating AGNs and galaxies at low redshift, when this color selection technique is applied to deeper samples, observations and templates suggest that a high degree of stellar contamination is unavoidable \citep{Donley2008}. The redshifts of our sample distribute at $2\leq z\leq3$ with much deeper IRAC data, we will identify AGNs from our sample with the mid-IR spectral index method next.

The slope of infrared SED of galaxy can be characterized by a power-law behavior of flux density with frequency $f_{\nu}\propto \nu^{\alpha}$. The infrared SED of AGN often follow a negative-sloping power-law, which may caused by either thermal or non-thermal emission originating the central region of galaxy. In contrast, stellar-dominated galaxies generally have positive IRAC power-law emission. For this reason, IRAC power-law selection has been proposed as a secure criteria for separating AGN-dominated candidates and normal galaxies with a small level of galaxy contamination \citep{Park2010,Donley2012,Fang2012}. In this work, the IRAC fluxes covering the 3.6 to 8.0 $\mu$m are fitted with a power-law $\alpha$ for each galaxy, and the $\alpha$ value is accepted only if the $\chi^2$ probability fit has $P_{\chi^2}>0.1$. Following \cite{Donley2008} and \cite{Park2010}, a limit of $\alpha \leq -0.5$ is chosen to classify galaxies as AGNs. We find 51 (19\%) eSFGs, 14 (24\%) cSFGs and 6 (12\%) cQGs can be selected as AGNs by this criterion, which also reveals that AGNs are more likely to be found in cSFGs.

We also match our sample with the $Chandra$ 1.8~Ms X-ray catalog in the COSMOS field \citep{Civano2012}, and find 2 eSFGs, 11 cSFGs and 3 cQGs have X-ray detections with $L_{0.5-10 \rm~keV}>10^{41}$ erg s$^{-1}$, as marked by cross symbols in Figure 7.
 The X-ray detected counterparts are more frequent among cSFGs than that in eSFGs and cQGs,
implying that cSFGs have experienced violent gas-rich interactions before, which could trigger both star formation
and black hole growth in an active phase.
Based on the $Chandra$ 4~Ms catalog from GOODS-S field, \cite{Barro2014a} found that about 47\% of all cSFGs at $2<z<3$ host an
X-ray-detected AGN. The AGN frequency in cSFGs of \cite{Barro2014a} is higher than those in our work. Comparison with the
$Chandra$ 4~Ms catalog in the GOODS-S field ($1.0\times10^{-17}\rm erg~cm^{-2}~s^{-1}$), the flux limit reached
in the COSMOS field (1.8 Ms) is $1.9\times10^{-16}\rm erg~cm^{-2}~s^{-1}$ in the full band ($0.5-10\rm~keV$).
Owing to only the most luminous AGNs can be detected at $z>2$, thus the intrinsic number could be higher if we were able to
detect lower-luminosity AGNs.

\begin{figure}
\includegraphics[width=1\columnwidth]{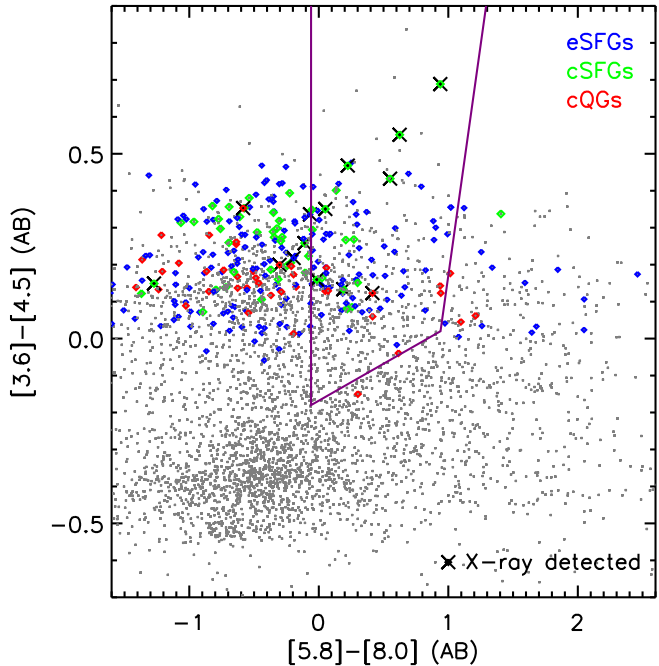}
\caption{IRAC mid-IR color-color diagram for galaxies in the COSMOS-CANDELS field. The solid purple lines indicate the boundaries of the AGN selection region introduced by Stern et al. (2005). Small points represent all galaxies with IRAC measurements in this field, eSFGs, cSFGs and cQGs are shown by blue, green and red squares, respectively. X-ray detected sources ($L_{0.5-10 \rm keV}>10^{41}$ erg s$^{-1}$) are marked by cross symbols.}
\end{figure}

\section{CONCLUSIONS}

In this work, we have described the construction of a large cSFG sample in the COSMOS-CANDELS field. Combined with the cSFGs selected from GOODS-S, we analyze the physical properties of cSFGs based on the multband photometry from 3D-HST and CANDELS imaging data.
Moreover, we also analyze our sample by using only the $z_{\rm UltraVISTA}$
to measure size and stellar mass, and we find that there is no change for the main results of cSFGs.
Our main conclusions are as follows:

1. The cSFGs are distributed at nearly the same locus on the MS as eSFGs, but they dominate the high-mass end and retain their highly active star-formation status into a compact phase. If we assume that cSFGs are the descendants of higher redshift eSFGs, they must have experienced gas-rich dissipative interactions, and enough amount of gas should also be remain in cSFGs to contribute to the high level of star-formation activity. We find no starbursts in our sample, indicating the starburst phase is short-lived and occurs anterior to the compact phase.

2. Most of the cSFGs have redder rest-frame $V-J$ colors and are closer to the quiescent region on the rest-frame $UVJ$ diagram. We find 20\% of the cSFGs located in the quiescent region, but they are closer to the boundary than are cQGs. At the same time, the cSFGs are distributed between eSFGs and cQGs in rest-frame $U-B$ color, and most of the cSFGs are located in the RS and GV regions on the rest-frame $U-B$ vs. $M_{\rm B}$ diagram. The results from the rest-frame $UVJ$ and rest-frame $U-B$ vs. $M_{\rm B}$ diagrams imply that a cSFG is a possible transitional phase of galaxy formation and evolution in terms of their stellar population properties.

3. We find that the cSFGs are distributed at the same locus as the cQGs on the $Gini$ vs. $M_{20}$ panel, but they are obviously different from eSFGs.
The cQGs and cSFGs have larger $Gini$ and smaller $M_{20}$, while eSFGs have the reverse. We derived the mean ($M_{20}$, $Gini$) values for cSFGs
(COSMOS+GOODS-S) as ($-1.54\pm0.21$, $0.55\pm0.10$), whereas eSFGs are ($-1.38\pm0.29$, $0.43\pm0.09$).
From visual inspection, we find that about one-third of cSFGs obviously show signatures of violent interactions or postmergers, and almost none of the cSFGs can be recognized as disks, implying that the progenitors of cSFGs have experienced violent gas-rich interactions such as dissipative wet mergers, which is considered to be the dominant mechanism for shrinking the size of a galaxy.
Moreover, we find that those visually extended cSFGs all have lower $Gini$ coefficients ($Gini<0.4$), indicating that
the $Gini$ coefficient could be used to clean out noncompact galaxies in a sample of candidate cSFGs.

4. To reduce the influence of cosmic variance on number density, we calculated the number density of cSFGs by combining cSFGs
from the COSMOS and GOODS-S fields. We find that massive ($M_{*}\geq10^{10}M_{\odot}$) cSFGs have a comoving number density of $(1.0\pm0.1)\times10^{-4}~{\rm Mpc}^{-3}$ at $2<z<3$. The general distributions of cSFGs on different physical parameters (the stellar population parameters: stellar mass $M_{*}$, dust extinction $A_{\mathrm V}$, stellar population age of galaxy, and the structural parameters: effective radius $r_{\rm e}$, S\'{e}rsic index $n$, ellipticity $1-b/a$) are very similar to that of cQGs. The existence of the similarity in physical properties between cSFGs and cQGs confirms the supposition that the cSFGs are short-lived and will be rapidly quenched into the quiescent phase.

5. We use three different methods (IRAC color-color diagram, mid-IR spectral index, and matching X-ray counterparts) to count the fraction of AGNs in our sample. We find in each of these methods that the cSFGs have the highest proportion of AGNs compared to eSFGs and cQGs, indicating
that cSFGs have previously experienced violent gas-rich interactions (by merging or disk instability), which could trigger both star formation
and black hole growth in an active phase.

\section*{Acknowledgments}

We are grateful to R. Abraham for access to his morphology analysis code. This work is supported by the National Natural Science Foundation of China (NSFC, Nos. 11303002, 11225315, 1320101002, 11433005, and 11421303), the Specialized Research Fund for the Doctoral Program of Higher Education (SRFDP, No. 20123402110037), the Strategic Priority Research Program "The Emergence of Cosmological Structures" of the Chinese Academy of Sciences (No. XDB09000000), the Chinese National 973 Fundamental Science Programs (973 program) (2015CB857004), the Yunnan Applied Basic Research Projects (2014FB155) and the Open Research Program of Key Laboratory for Research in Galaxies and Cosmology, CAS.


\end{document}